\documentclass[sigconf]{acmart}
\usepackage{siunitx}
\AtBeginDocument{%
  \providecommand\BibTeX{{%
    \normalfont B\kern-0.5em{\scshape i\kern-0.25em b}\kern-0.8em\TeX}}}

\copyrightyear{2023} 
\acmYear{2023} 
\setcopyright{acmlicensed}
\acmConference[CSCW '23 Companion]{Computer Supported Cooperative Work and Social Computing}{October 14--18, 2023}{Minneapolis, MN, USA}
\acmBooktitle{Computer Supported Cooperative Work and Social Computing (CSCW '23 Companion), October 14--18, 2023, Minneapolis, MN, USA}
\acmPrice{15.00}
\acmDOI{10.1145/3584931.3606976}
\acmISBN{979-8-4007-0129-0/23/10}

\begin{document}
\title{VR Job Interview Using a Gender-Swapped Avatar}

\author{Jieun Kim}
\authornote{These authors contributed equally}
\email{jk2345@cornell.edu}
\orcid{0000-0002-1530-8871}
\affiliation{%
  \institution{Cornell University}
  \city{Ithaca}
  \state{NY}
  \country{USA}
  \postcode{14853}
}

\author{Hauke Sandhaus}
\authornotemark[1]
\email{hgs52@cornell.edu}
\orcid{0000-0002-4169-0197}
\affiliation{%
  \institution{Cornell University}
  \city{Ithaca}
  \state{NY}
  \country{USA}
  \postcode{14853}
}

\author{Susan R. Fussell}
\email{sfussell@cornell.edu}
\orcid{0000-0001-8980-5232}
\affiliation{%
  \institution{Cornell University}
  \city{Ithaca}
  \state{NY}
  \country{USA}
  \postcode{14853}
}

\renewcommand{\shortauthors}{Kim et al.}

\begin{abstract}

Virtual Reality (VR) has emerged as a potential solution for mitigating bias in a job interview by hiding the applicants' demographic features. The current study examines the use of a gender-swapped avatar in a virtual job interview that affects the applicants' perceptions and their performance evaluated by recruiters. With a mixed-method approach, we first conducted a lab experiment (N=8) exploring how using a gender-swapped avatar in a virtual job interview impacts perceived anxiety, confidence, competence, and ability to perform. Then, a semi-structured interview investigated the participants’ VR interview experiences using an avatar. Our findings suggest that using gender-swapped avatars may reduce the anxiety that job applicants will experience during the interview. Also, the affinity diagram produced seven key themes highlighting the advantages and limitations of VR as an interview platform. These findings contribute to the emerging field of VR-based recruitment and have practical implications for promoting diversity and inclusion in the hiring process.

\end{abstract}


\begin{teaserfigure}
    \centering
    \includegraphics[width=1\textwidth]{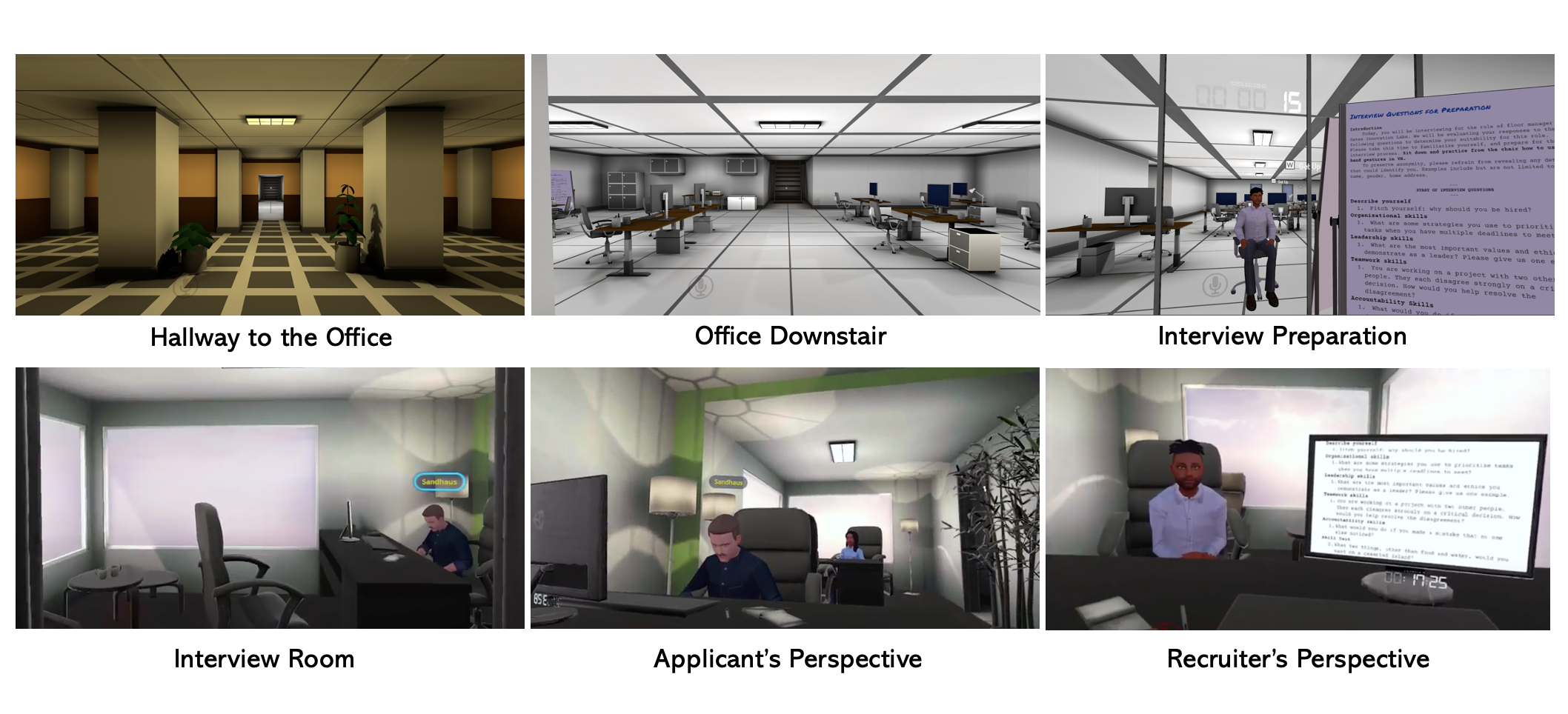}
    \caption{A two-story virtual office was created using Unity and VRchat, featuring realistic assets. The first floor has a mirror and whiteboard for interview preparation, while the second floor has an interview room for applicants and recruiters.}
    \label{fig1}
\end{teaserfigure}

\received{6 July 2023}
 \maketitle

\section{Background}

Virtual Reality (VR) is an emerging platform that allows users to engage with the simulated environment from a first-person perspective. As it offers rich and immersive experiences that closely resemble face-to-face interactions, organizations across the board (e.g., US Navy, General Mills, and Jaguar) are increasingly turning to VR for interviewing job applicants \cite{VRint}. Earlier studies viewed VR as a viable alternative to traditional interview settings. VR can help individuals to feel more focused and less distracted during interviews by controlling the external audio and visual cues irrelevant to the interview. On the other hand, the lack of nonverbal feedback in VR makes it difficult to discuss with the recruiter compared to face-to-face interviews \cite{beti2019efficacy}.

While VR can simulate the real-life environment, using gender-stereotyped avatars can perpetuate the existing biases in the hiring process. Over the decades, gender has heavily influenced recruitment practices, leading to gender bias where one gender is perceived as more competent than the others \cite{bailey2019man}. Studies have shown that even when male and female candidates possess similar qualifications and experiences, recruiters consistently rate male candidates as more competent and hireable \cite{moss2012science, cheong2020ethical}. This bias also affects female applicants' chances of being selected for faculty \cite{steinpreis1999impact, nittrouer2018gender} and industry positions \cite{funk2018women}. 

We propose the design of gender-swapped avatars, which represent the gender incongruent with the users' self-identified gender, as the possible solution to address cognitive bias revealed by both applicants and recruiters in job interviews. Using gender-swapped avatars in the virtual interview can conceal the job applicants' physical identity, which can help mitigate (un)conscious bias in the hiring process. A recent study tested if using a gender-swapped avatar helps reduce implicit bias and increase empathy toward women in the STEM field \cite{crone2022interview}. The study found that male participants using a gender-swapped avatar (i.e., female avatars) chose the female candidate more often compared to when they were using a gender-congruent avatar (i.e., male avatars). This highlights the users’ ownership over the virtual body which affects their perceptions and actual behaviors in the virtual space.

As VR is becoming a potential venue for job interviews, it is essential to understand how the applicants' avatar representation affects users’ perceptions and behaviors. As the first step to designing a gender-inclusive interview environment in VR, the current study investigates the applicants’ psychological responses when manipulating the avatars' gender which may affect their interview performance. With eight university senior students who prepare job applications and interviews, this study conducted a preliminary experiment investigating the impact of the avatar’s gender on VR experiences, followed by a semi-structured interview that discovers the advantages and disadvantages of VR as an interview platform. Employing a mixed-method approach, we explore three research questions: 

\begin{itemize}
    \item{\textit{\textbf{RQ1.} How does the applicants’ use of gender-swapped avatars affect their perceptions regarding (a) anxiety, (b) confidence, (c) competence, and (d) ability to perform?}}
    \item{\textit{\textbf{RQ2.} How does the applicants’ use of gender-swapped avatars affect the recruiters’ evaluation of the applicant’s perceived (a) anxiety, (b) confidence, (c) competence, and (d) ability to perform?}}
    \item{\textit{\textbf{RQ3.} How does using an avatar in VR affect the applicants' interview experiences?}}
\end{itemize}

Our study revealed that participants using a gender-swapped avatar reported a significantly lower level of anxiety compared to those using a gender-matched avatar. Additionally, participants reported that using avatars during the virtual interview helped alleviate their concerns about being judged on their appearance, which enabled them to feel more comfortable and focus on their interview questions. By shedding light on the potential benefits and challenges of using VR in job interviews, our study can provide valuable insights to researchers and practitioners in the CSCW community for designing virtual job interviews.

\section{Method}

\subsection{Overview}
To answer three research questions, we first operated a lab-based between-subject experiment where participants conduct a simulated job interview in the VR using either gender-swapped or gender-matched avatars. The avatar's gender was manipulated by either swapping or aligning its visual features with the participant's self-identified gender (See details in Materials). After conducting the VR interview, the participants evaluated their interview experience in terms of their perceived anxiety, confidence, competence, and ability to perform (RQ1). They also watched the other applicants' recorded interview videos taking the role of recruiter to assess their perceived attitudes (RQ2). Following that, a semi-structured interview was conducted to gather insights on the participants' experience of using VR and avatars for job interviews (RQ3).

\subsection{Participants}
The university students who are looking for or preparing for job interviews (age; \textit{M}=23.28, \textit{SD}=4.02) were recruited from a university recruitment system that grants course credits for participation. Participants (\textit{N}=8) were made up of five males and three females, and none of them were identified as transgender or non-binary gender. In the study, participants wore head-mounted devices (i.e., Oculus Quest 2) and used hand controllers to navigate the VR space. In terms of VR experience, one had “never” used it, two had “rarely” used it, two used it “sometimes”, and one used it “often”. The remaining two participants did not indicate their VR experience.

\begin{figure}
    \centering
    \includegraphics[width=0.45\textwidth]{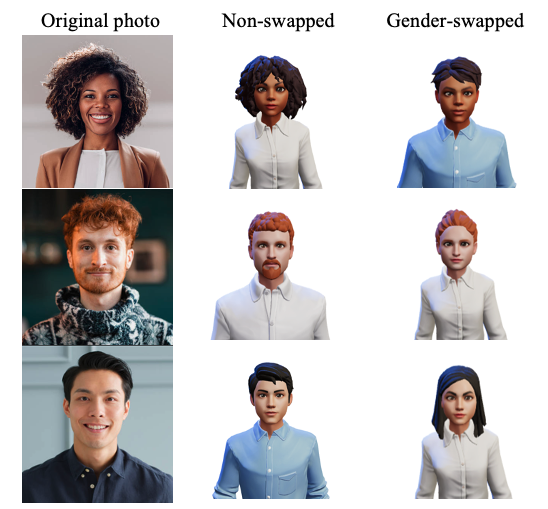}
    \caption{Examples of VR avatars}
    \label{fig2}
\end{figure}

\subsection{Materials}

\subsubsection{Virtual office in VRchat}
Using the Unity game engine and the VRchat SDK, we built a two-story virtual office (Fig.~\ref{fig1}). We included decorative assets from the Unity asset store, such as coffee mugs, printers, notepads, and high-quality lights to make the office look realistic. On the first floor, the major interactive component in the office is a large-scale mirror next to a whiteboard that displays the interview questions. By putting a whiteboard in front of the mirror, we intentionally led the participants to see themselves in the mirror often while preparing for the interview, which will strengthen the manipulation of avatar embodiment. On the second floor, connected via a stair, the recruiter's office was furnished with chairs where applicants can sit in front of the recruiters during the interview.

\subsubsection{Avatar and voice manipulation}
The Ready Player Me platform allows the automatic creation of avatars with the users’ face images. This automated process eliminates potential biases from researchers when customizing participants' avatars. The platform simulates the avatar's visual features such as race, age, hair color, face shape, and gender to align with the user's characteristics, while allowing the researcher to modify or keep the assigned gender as per the experimental conditions. To minimize confounding factors unrelated to gender, the avatars' attractiveness and likeability were controlled through a pretest. Fig.~\ref{fig2} shows examples of automatically generated avatars from profile photos \cite{image1, image2, image3} and gender-swapped avatars. Their full-body avatars simulate eye blinking, individual finger movements, idle body movements, and animate mouth movements. The voice of participants in the gender-swap conditions was manipulated through MorphVOX pro by increasing or lowering their pitch. A voice anonymization filter obfuscated all participants' voices.

\subsubsection{Job interview questions} 
The job interview questions targeted ‘soft skills’ (e.g., organizational skills, leadership skills, teamwork skills, accountability skills) which are not associated with the participant’s particular job expertise or capabilities. To control the participants' preferences or perceived difficulties, those questions did not require them to reveal personal information and explicit knowledge for the right answers. The questions were refined in several sessions to control the confounding effects.

\subsection{Procedure}

\begin{figure}
    \centering
    \includegraphics[width=0.5\textwidth]{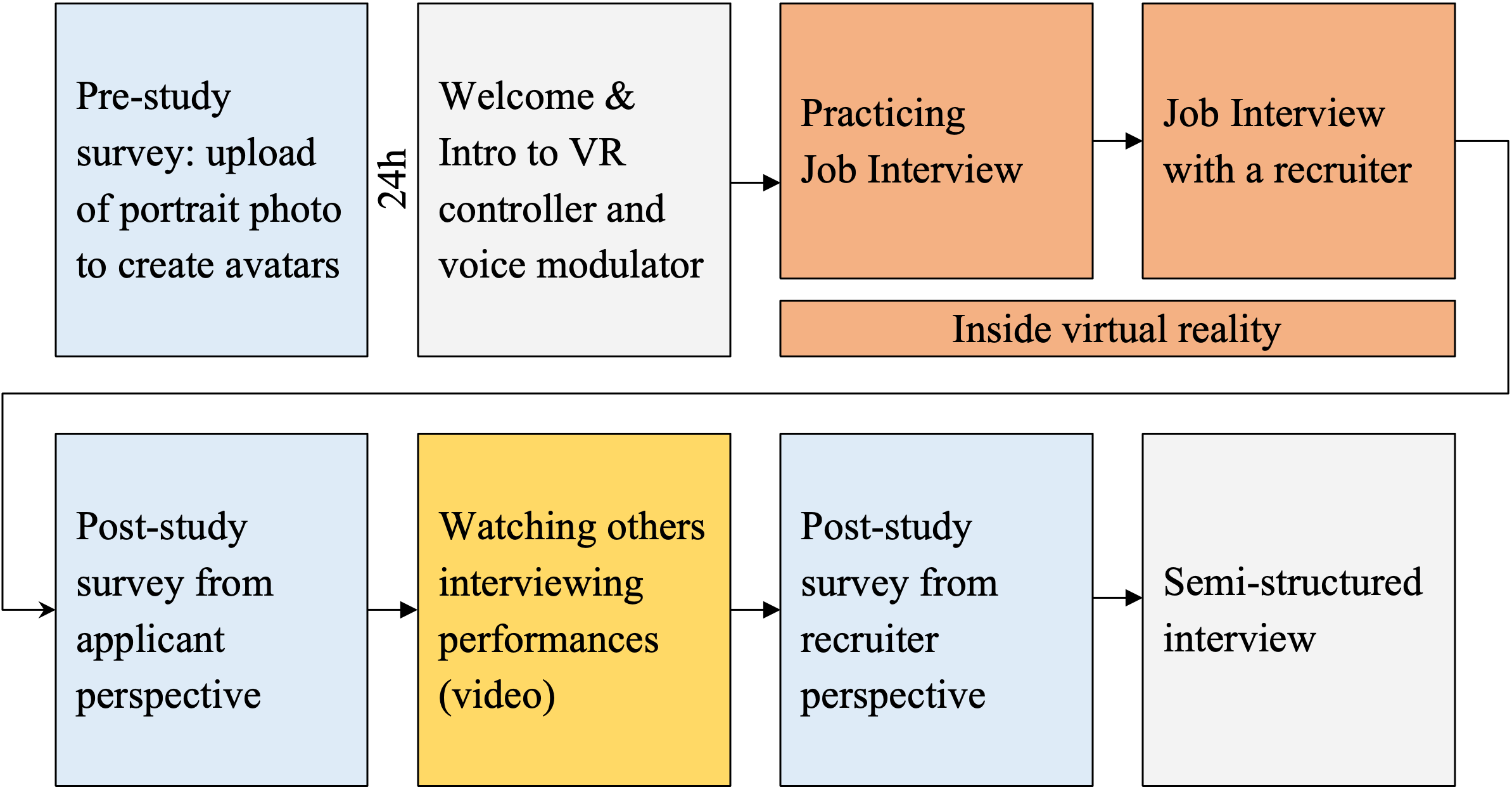}
    \caption{Study procedure}
    \label{fig4}
\end{figure}

Fig.~\ref{fig4} summarizes the procedure that participants go through during the study. Before starting the VR interview simulation, participants were asked to submit their face photos to generate their avatars used in the VR space. The gender of each avatar was manipulated to be either congruent or incongruent with their self-identified gender. The platform also supports non-gendered avatar bodies, but in our sample, all participants self-identified with one of the binary genders. After being informed about how to use VR controllers and voice modulators, participants entered the virtual space and were asked to spend five minutes in front of a mirror to familiarize themselves with the avatar's appearance, gestures, and body movements. Also, they got additional five minutes to prepare the interview questions written on a virtual whiteboard by looking at themselves in the mirror. Then, they were instructed to go upstairs to join the recruiter in the interview room while maintaining their anonymity. The recruiter avatar was represented as a white male controlled by a confederate researcher. The recruiter followed a pre-written script and only provided standardized listening feedback. Participants answered five interview questions which were recorded via the confederate's headset.

After exiting the virtual office, participants completed a post-survey to evaluate their anxiety, confidence, competence, and ability to perform. After that, participants watched two interview recordings of other participants randomly selected across the conditions and evaluated their interview performance regarding perceived anxiety, confidence, competence, and ability to perform. Finally, participants were asked about the overall experience of VR interviews, conducting a semi-structured interview for about 10 minutes.

\subsection{Measures}

\subsubsection{Post-test survey}
The post-survey questionnaires were used both when the participants evaluate their own perceptions on VR interview and when they evaluate other participants’ interview performance after watching the recordings. The measures include anxiety \cite{mcfarland2004examination}, confidence \cite{carver1994behavioral}, competence, and ability to perform \cite{bauer2001applicant} retrieved and adapted from the questionnaires of previous studies, using a 5-point Likert scale (1: Strongly disagree - 5: Strongly agree). 

\subsubsection{Semi-structured interview}
We used open-ended questions for a semi-structured interview. These questions probe the following topics: specifics about VR experience, virtual reality versus face-to-face for a job interview, the influence of the avatar, and emotions experienced during the experiment. After recording all participants' post interviews answers, we transcribed them with AI software, followed by manual correction by two researchers. To find meaningful themes from the data, we used affinity diagramming to group excerpts from those transcripts into clusters and analyzed relationships between them (Appendix. Fig.~\ref{fig3}). All clusters had mentions from more than half of the participants.

\section{Results}

\renewcommand{\arraystretch}{1.05}
\begin{table}
\centering
\caption{Mann-Whitney U Test: VR job applicants' self-evaluation}
\label{table}
\resizebox{\columnwidth}{!}{%
\begin{tabular}{l *{8}{S[table-format=2.2]}}
\toprule
& \multicolumn{2}{c}{Gender-matched (n=4)} & \multicolumn{2}{c}{Gender-swapped (n=4)} & \multicolumn{3}{c}{Test Statistics} \\ 
\cmidrule(lr){2-3} \cmidrule(lr){4-5} \cmidrule(lr){6-8}
& {\textit{Mean Rank}} & {\textit{Sum of Ranks}} & {\textit{Mean Rank}} & {\textit{Sum of Ranks}} & {\textit{U}} & {\textit{Z}} & {\textit{p}} \\ \midrule
Anxiety & 7.00 & 21.00 & 3.00 & 15.00 & .000 & -2.23 & .02* \\ 
Confidence & 4.67 & 14.00 & 4.40 & 22.00 & 7.00 & -0.15 & .87 \\
Competence & 5.67 & 17.00 & 3.80 & 19.00 & 4.00 & -1.05 & .29 \\
Ability to perform & 4.33 & 13.00 & 4.60 & 23.00 & 7.00 & -0.15 & .87 \\ \bottomrule
\end{tabular}%
}
\footnotesize{\textit{Note.} *asymptotic sig.}
\end{table}

\subsection{Quantitative data analysis}
Given the small sample size (\textit{N}=8), we employed the Mann-Whitney U test in our study. The Mann-Whitney U test is used to compare differences between two independent groups that are not normally distributed \cite{mann1947test}. It has the great advantage of being used for small sample sizes of subjects, such as those with less than 15 participants \cite{nachar2008mann}, while producing significant results comparable to those of the t-test \cite{zimmerman1987comparative}. Table.~\ref{table} presents the Mean Rank and Sum of Ranks for each condition group along with the statistical significance of the participants' perceptions during the VR job interview. Our findings concerning RQ1 indicate that participants who used gender-swapped avatars tended to experience significantly lower anxiety than those who used gender-matched avatars, \textit{z}=-2.23, \textit{p}=.02. However, there were no significant differences in perceived confidence, competence, and ability to perform between the two groups. Regarding RQ2, the avatar did not significantly affect how participants evaluated other participants in terms of perceived anxiety, \textit{z}=-.77, \textit{p}=.43, confidence, \textit{z}=-1,17 \textit{p}=.24, competence, \textit{z}=-1.85, \textit{p}=.06, and ability to perform, \textit{z}=-.47, \textit{p}=.63.

\subsection{Qualitative data analysis} 
Based on the interview responses, we identified 7 major clusters: (1) The main advantage of VR interviews people described was the ability to be assessed based on skills and qualifications, rather than appearance. Using an avatar allowed the users to remain good-looking and anonymous about their personality traits, such as gender, race, and age. Interestingly, some participants assumed that the recruiters would also be unaffected by the visual appearance of the avatar, as they would be aware that anyone can be anyone behind it. (2) Many people found the virtual interview environment to be immersive and similar to real-life interviews. (3) Anonymity and not being there in person made people feel less stressed and more comfortable than in face-to-face interviews. People who previously did job interviews preferred VR over face-to-face interviews. Some people also believed that VR could be a useful training platform. (4) However, some people struggled to adapt to the virtual environment and experienced motion sickness during prolonged use. (5) The lack of facial expressions in avatars discouraged some participants, as limited feedback from the recruiter avatar prevents them from gauging how the interviewer evaluates the applicant during the interview. (6) Participants highlighted the importance of gestures in establishing a connection with others and feeling present in VR. While some believed that the limited gestures might impact their performance negatively, others assumed that recruiters would take this limitation into account when evaluating the applicants. (7) Lastly, participants emphasized the importance of customizing their avatars based on their self-identity in order to feel present and immersed in the virtual environment. They wanted to dress their avatars appropriately for the job interview, and none of them wanted to create an avatar that did not match their physical appearance.

\section{Discussion and limitations}

Our findings show the trend that participants using gender-swapped avatars experience a lower level of anxiety than those using gender-matched avatars. Considering the results from our qualitative data analysis, we can assume that gender-swapped avatars might lead the participants to feel less anxious under the umbrella of anonymity. Using an avatar that less represents themselves might reduce their pressure under the interview setting. Another possibility is that gender-swapped individuals might feel less connected to their virtual environment and their avatars, which may lower their pressure to perform well. Therefore, the avatar's gender swap may have alleviated the anxiety they might have felt to act better in the interview. Moreover, the avatar's gender swap did not affect how the recruiter evaluate the applicants' attitudes and performance. This could prove beneficial in practice, demonstrating that even when the interview applicants are represented as a different gender, it does not make significant bias in the recruiting process.

The qualitative interview data proposed the advantages and limitations of VR as an interview platform. Consistent with the findings reported in previous qualitative research \cite{beti2019efficacy}, users who conducted VR interviews felt comfortable and less conscious, enabling them to focus on their interview answers. While most of the participants were impressed by how similar the virtual interview space was to the real-life setting, still VR devices have limited capacity to simulate participants' real movements on a delicate level, such as a lack of nonverbal cues. Some participants mentioned that gestures are critical to providing natural and present interactions with the recruiter avatar. Also, VR limits understanding of the recruiter avatar's facial expression which indicates how the recruiter thinks about the applicants and facilitates VR users to make realistic and effective interactions. The technical limitations of VR contributing to motion sickness need to be addressed to create a comfortable and immersive VR experience for users

As recruiting process is the first stage of an individual's professional career, initial biases formed during this process can cause pay disparities and other long-term effects along the road. Social constructs may lead candidates to believe that they are already at a disadvantage because of their social identities. In addition, a job interview tends to be a stressful setting, which can also hinder the candidate's performance. In this sense, virtual reality can be a promising tool for conducting job interviews once the technical limitations are overcome. As the study provides preliminary results with a small sample size, we could not identify gender biases in the job performance evaluation process. To identify these biases, the next step will be to expand the participation pool to a larger population that is not skewed toward university students. Another limitation of this study is that the participant did not include diverse gender groups (e.g., transgender and non-binary gender). Including different gender groups that are non-binary will provide another aspect of using gender-swapped avatars that varies users' interview experience. Future studies are required to examine how the embodiment of avatars that presents users' identities such as race, gender, age, and physical traits can form more inclusive experiences for diverse groups.

\bibliographystyle{ACM-Reference-Format}
\bibliography{reference.bib}

\onecolumn
\appendix
\counterwithin{figure}{section}

\section{Appendix}
    \begin{figure}[h]
    \centering
    \includegraphics[width=1\textwidth]{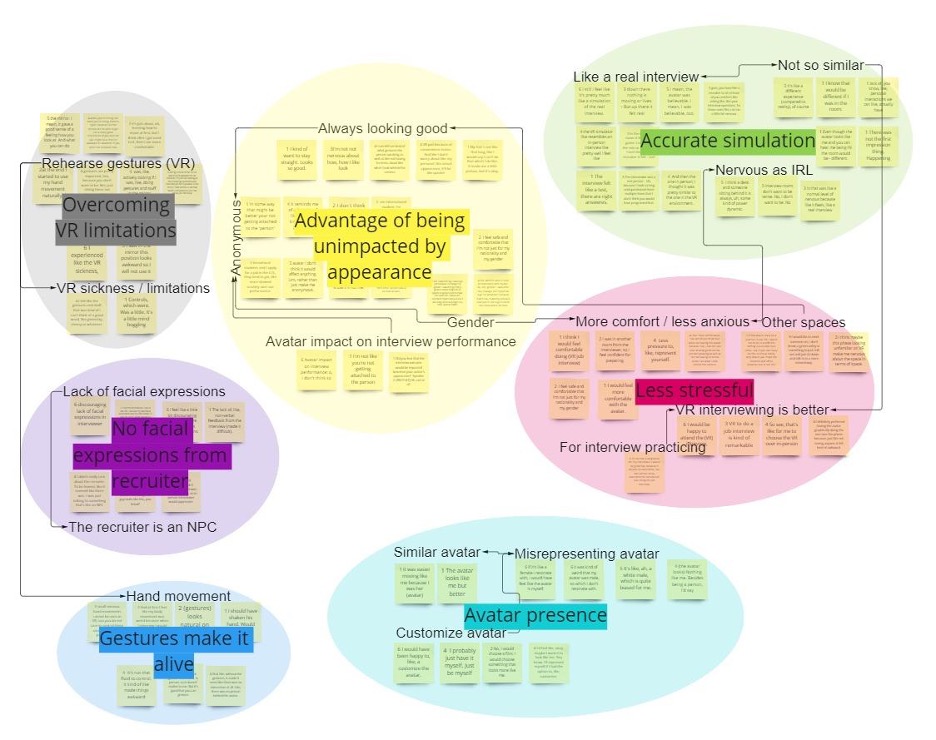}
    \caption{Affinity Diagram}
    \label{fig3}
    \end{figure}
    
\end{document}